\documentclass[onecolumn]{elsart}
\usepackage{natbib}
\usepackage{graphicx}
\begin{document}
\runauthor{Spatially resolved host of GRB 060505}
\begin{frontmatter}
\title{The spatially resolved host of GRB 060505 and implications for the nature of the progenitor}
\author[Dark]{Christina C. Th\"one},
\author[Dark]{Johan P. U. Fynbo}

\address[Dark]{Dark Cosmology Centre, Niels Bohr Institute, University of Copenhagen, Juliane Maries Vej 30, 2100 Copenhagen, Denmark}
\begin{abstract}
We present a study of the host galaxy of the Gamma-Ray Burst (GRB) of May 5
2006 (GRB\,060505). The host is spatially resolved in both imaging data and in
a long slit spectrum including the GRB site. We find the galaxy to be a Sbc
spiral, which is unusual for a long GRB host galaxy. The site of the GRB
is considerably different from the rest of the galaxy with intense star 
formation, low metallicity and a young age. This suggest a massive stellar
progenitor rather than a merger of compact objects which has been suggested 
based on the the relatively short duration of T$_{90}$=4s for the prompt
emission.
\end{abstract}
\begin{keyword}
GRBs: GRB 060505, GRB host galaxies, emission lines
\end{keyword}
\end{frontmatter}

\section{Introduction}
Until May 2006, the division of GRBs in short  (T$_{90}$ $<2s$) and long
duration bursts (T$_{90}$ $>$2s) and the connection with different progenitors
gained increasing evidence. Short GRBs should originate from the merging of two
compact objects and be found mostly in old populations and possibly large
distances from their birth site due to kicks from the supernova explosion. Long
duration bursts have been firmly linked to supernovae (SNe) \cite{Hjorth03} and
had been almost always found in young stellar populations and star forming
galaxies \cite{Fruchter}.\\
With two nearby long-duration GRBs which lack any sign of a SN in their lightcurves, GRB\,060505 and GRB\,060614 \cite{Gehrels06, Fynbo06, DellaValle06, Gal-Yam06}, this picture has been challenged. Suggestions have been brought up whether this can be explained by extinction, the faintness of the SN or if a new type of GRB progenitor is needed. In the case of GRB\,060505, its T$_{90}$ of only 4s could also suggest a compact merger origin.\\
One way to reveal the nature of the GRB progenitor is to investigate the properties of the burst site. Due to their distance, we are usually restricted to analyze the global properties of GRB host galaxies, but as most long GRB hosts are relatively small, the properties of the host and the environment are likely to be comparable. This is however different for large hosts such as the spiral host of GRB\,060505, whose properties can vary significantly over the galaxy. This we investigate in this paper by analyzing a spatially resolved spectrum of the host.

\section{Dataset}
We used FORS2 at the VLT and grism 300V with a 1.0'' slit (resolution: 11 \AA{}) on May 23 to obtain a series of longslit spectra of the host galaxy which were stacked into a single spectrum. From the spectrum covering the bulge of the galaxy, the GRB site in the northern arm and a spiral arm south of the bulge, five traces from the different regions were extracted and flux calibrated. Furthermore, we obtained FORS1+ISAAC/VLT images in UBVRIzK taken on Sep. 14, 24 and Oct. 1. Further details on the reduction and calibration of the spectra and the images can be found in \cite{Thoene07}.

\section{The nature of the host galaxy}
The host galaxy of GRB\,060505 is classified as an Sbc spiral from the morphology, the strength of the emission lines and the colors. Spiral galaxies are the exceptions among long GRB host galaxies, there are however some earlier examples, namely GRB\,980425 \cite{Sollerman05}, GRB\,990705 \cite{LeFloch02}, GRB\,020819 \cite{Jakobsson} and possibly GRB\,051109BÊ \cite{Perley06}, of which one, GRB\,980425 was clearly  connected to a SN \cite{Galama}. Remarkably, the GRB has been found in the outer parts of their hosts in all these cases.\\
For GRB\,060505 it was possible for the first time to derive a rotation curve for a GRB host galaxy by tracing the Doppler shift of four strong emission lines
along the spatial profile. The rotation curve flattens at a velocity of 212 km/s, taking into account the inclination of the galaxy of 49 deg. From that we derive a mass within the half-light diameter (24 kpc) of 1.3$\times$ 10$^{11}$ M$_\odot$.\\

\section{The GRB site}
\subsection{Burst location and extinction}
GRB\,060505 was situated in the northern spiral arm of its host galaxy within a bright star-forming region \cite{Fynbo06} as confirmed also by HST imaging \cite{Ofek07}. This speaks in favor of the collapsar origin as short GRBs are expected to be found far away from the birth sites of their compact object progenitors. \cite{Ofek07} calculated a minimum merger age of 10 Myr from the size of the star-forming region assuming a very low kick velocity. This is on the lower limit of the delay time for a compact object merger \cite{Belczynski}.\\
The extinction measured from both the afterglow spectral energy distribution (Xu et al., in prep.) and the spectrum of the burst site show a very low extinction of A$_V$$<$ 0.09 mag. This again rules out the possibility that the SN was simply missed due to extinction.

\subsection{Metallicity and SFR}

\begin{figure*}
\begin{center}
\begin{minipage}{9cm}
\vspace{3mm}
\includegraphics[width=9cm]{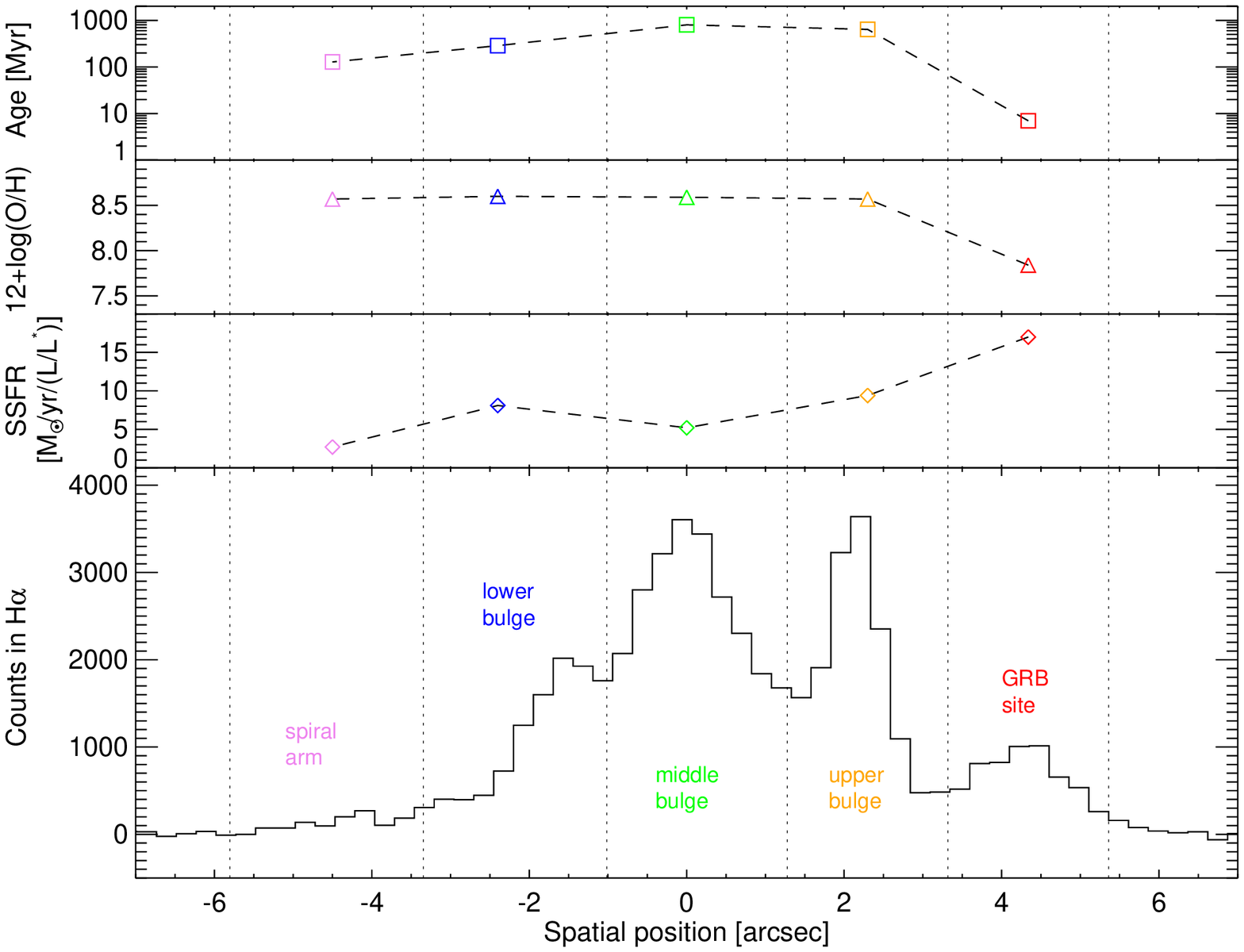}
\end{minipage}
\begin{minipage}{4.5cm}
\includegraphics[width=4cm]{f1.eps}\\[0.5mm]
\includegraphics[width=4cm]{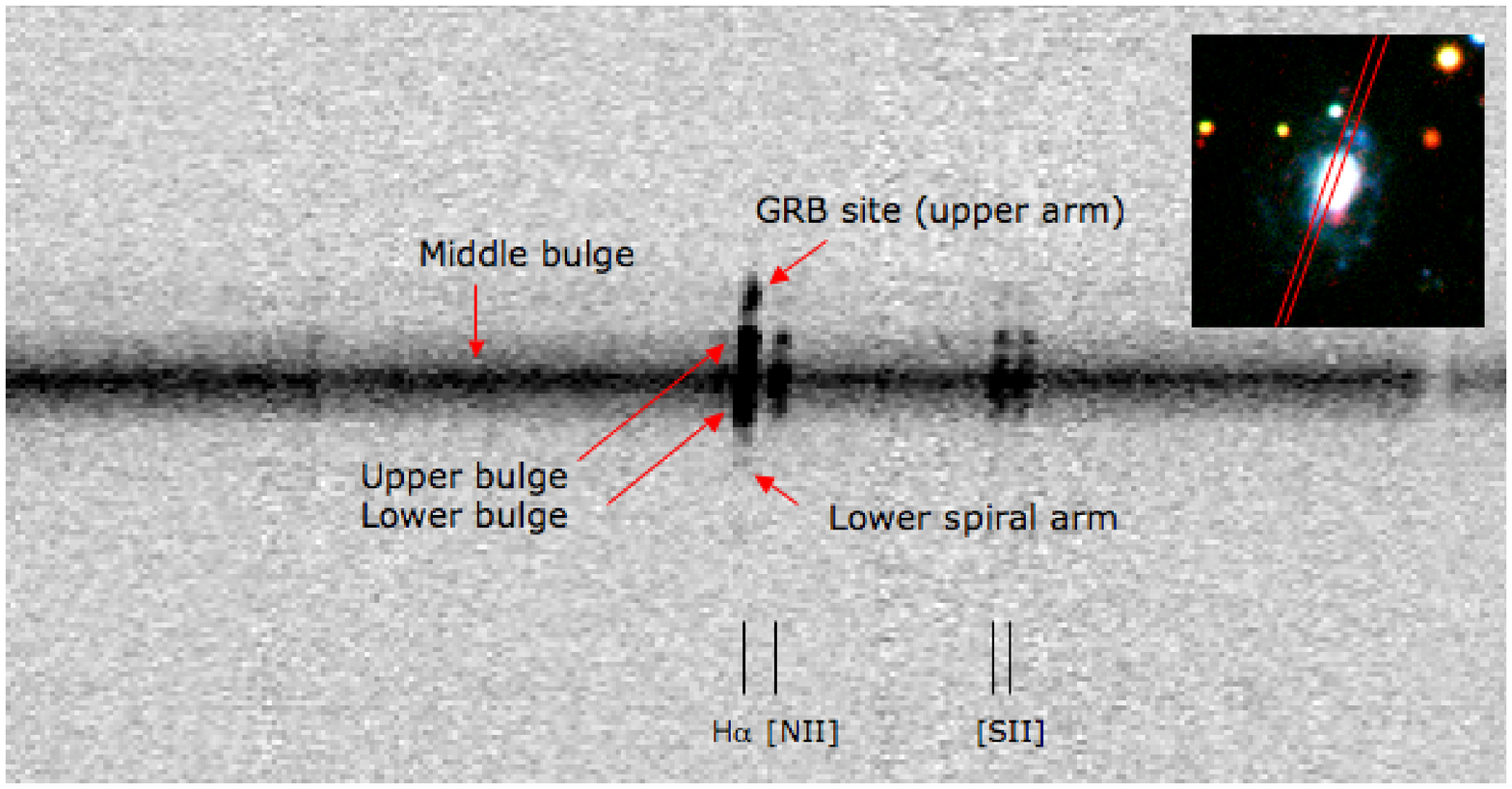}
\end{minipage}
\end{center}
\caption{Properties of the host galaxy in 5 different parts of the host galaxy, indicated in the 2D spectrum. For the full 2D spectrum see also \cite{Thoene07,Watson07}}
\label{figure}
\end{figure*}

An important ingredient in the modelling of collapsars that can produce a GRB is the metallicity of the progenitor star. Only metal poor stars are believed to preserve enough angular momentum to be able to lauch a jet \cite[e.g.]{xx}. Long GRBs have genereally been found in sub-solar metallicity environments \cite[e.g.]{xy} which supports this theory.\\
We determined the metallicity in the five different spectra from the Oxygen emission lines using the so-called R$_{23}$ parameter \cite{Pagel}, which has been widely used for other GRB hosts. We further apply the latest recalibration from \cite{Kewley}. This tentativlely gives a very low Oxygen abundance of log(12+O/H)=7.8 for the GRB site whereas it is around or even above solar in the rest of the galaxy (see Fig.~\ref{figure}).\\
From the H$\alpha$ flux which is proportional to the amount of young stars, we derive the star-formation rate (SFR) in the different parts of the galaxy and scale it with the V band luminosity in the same parts. We then find a moderate specific SFR of 3 to 8 M$_\odot$/yr/(L/L*) in the bulge and a high SF of 18 M$_\odot$/yr/(L/L*) at the GRB region. Both the low metallicity and the high SFR at the GRB site give an indication for a collapsar origin of the GRB rather than a merger event. 

\subsection{A very young age for the stellar population.}
We used several different approaches for determining the age of the stellar population in the different parts of the galaxy. The GRB site is a star-forming HII region which speaks for a low age of the stellar population. For very young populations, the EW of H$\alpha$ gives an upper limit on the age of the population \cite{Zackrisson02}. Our H$\alpha$-EW of $-181$ \AA{} leads to an age of $<$ 7 Myr which basically excludes the possibility of a merger event even for the shortest merger timescales \cite{Belczynski}. For the other parts of the galaxy, we used stellar population models from Bruzual\&Charlot \cite{BruzualCharlot} and find ages that are all above 10$^8$ yr (see also Fig. 1).

\section{Conclusion}

The comparison of the GRB site with the rest of the host galaxy reveals a substantial difference in its properties. Whereas the host galaxy itself turns out to be a regular late-type Sbc spiral, the GRB site has properties close to those of the more common type of long GRB host galaxy, low-mass, low metallicity, highly starforming dwarf galaxies. The position of the GRB right within a star forming region and the very young age of the stellar population speaks much more in favour of a collapsar progenitor than a merger of two compact objects. The low metallicity is a further hint for the massive star progenitor while a merger could occure in any environment. \cite{Fruchter} had proposed that if long GRBs occure in spirals, they should be found in the outer spiral arms in younger populations and low metallicity environments as we see here. We conclude that this GRB was very likely coming from the death of a massive star rather than a merger and the prognitor might just not have produced a (bright) supernova.

{\bf Acknowledgements}\\
\ \\
The ``Dark Cosmology Centre'' is funded by the DNRF.


\begin{thebibliography}{999}

\bibitem{Hjorth03} Hjorth, J. et al. 2003, Nature 423, 847

\bibitem{Fruchter} Fruchter, A. S. et al. 2006, Nature, 441, 463

\bibitem{Gehrels06} Gehrels, N. et al. 2006, Nature 444, 1044

\bibitem{Fynbo06} Fynbo, J. P. U. et al 2006, Nature 444, 1047

\bibitem{DellaValle06} Della Valle, M. et al. 2006, Nature 444, 1050

\bibitem{Gal-Yam06} Gal-Yam, A. et al. 2006, Nature 444, 1053

\bibitem{Thoene07} Th\"one, C. C. et al. 2007, ApJ submitted, preprint: astro-ph/0703407

\bibitem{Sollerman05} Sollerman, J. et al. 2005, New Astronomy 11, 103

\bibitem{LeFloch02} Le FlocÕh, E. et al. 2002, ApJ, 581, L81

\bibitem{Jakobsson} Jakobsson, P. et al. 2005, ApJ, 629, 45

\bibitem{Perley06} Perley, D. et al. 2006, GCN $\#$ 5387

\bibitem{Galama} Galama, T. J. et al. 1998, Nature 395, 670

\bibitem{Ofek07} Ofek, E. et al. 2007, ApJ accepted, astro-ph/0703192

\bibitem{Belczynski} Belczynski, K. et al. 2006, ApJ 648, 1110

\bibitem{xx} Hirschi, R. et al. 2005, A\&A, 443, 581

\bibitem{xy} Fynbo, J. P. U. et al. 2006, A\&A, 451, L47

\bibitem{Pagel} Pagel, B. E. J. et al. 1979, MNRAS 189, 95

\bibitem{Kewley} Kewley, L. J. et al. 2007, AJ 133, 882

\bibitem{Zackrisson02} Zackrisson, E. et al. 2001, A\&A, 375, 814

\bibitem{BruzualCharlot} Bruzual, G. \& Charlot, S. 2003, MNRAS, 344, 1000

\bibitem{Watson07} Watson, D. et al. 2007, PTRSA, 365, 1269




\end{thebibliography}
\end{document}